\documentclass{article}

\PassOptionsToPackage{numbers, compress}{natbib}

\usepackage[preprint]{neurips_2024}

\usepackage[utf8]{inputenc} 
\usepackage[T1]{fontenc}    
\usepackage{hyperref}       
\usepackage{url}            
\usepackage{booktabs}       
\usepackage{amsfonts}       
\usepackage{nicefrac}       
\usepackage{microtype}      
\usepackage{xcolor}         
\usepackage{graphicx} 
\usepackage{multirow} 
\usepackage{xspace}
\usepackage{wrapfig}
\usepackage{enumitem}
\setlist[itemize]{leftmargin=*}

\usepackage{algorithm}
\usepackage{algpseudocode}

\newcommand{\rankfusion}{\texttt{LLM-RankFusion}\xspace}
\newcommand{\relzero}{{\color{magenta}\textbf{0}}\xspace}
\newcommand{\relone}{{\color{orange}\textbf{1}}\xspace}
\newcommand{\reltwo}{{\color{blue}\textbf{2}}\xspace}
\newcommand{\relthree}{{\color{teal}\textbf{3}}\xspace}
\newcommand{\relzerop}[1]{{\color{magenta}\textbf{#1}}\xspace}
\newcommand{\relonep}[1]{{\color{orange}\textbf{#1}}\xspace}
\newcommand{\reltwop}[1]{{\color{blue}\textbf{#1}}\xspace}
\newcommand{\relthreep}[1]{{\color{teal}\textbf{#1}}\xspace}

\title{LLM-RankFusion: Mitigating Intrinsic Inconsistency in LLM-based Ranking}

\author{
Yifan Zeng\textsuperscript{1,*}, Ojas Tendolkar\textsuperscript{1,*}, Raymond Baartmans \textsuperscript{1,*}, \\
\textbf{Qingyun Wu\textsuperscript{2}}, \textbf{Lizhong Chen\textsuperscript{1}}, \textbf{Huazheng Wang\textsuperscript{1}} \\
\textsuperscript{1}Oregon State University, \textsuperscript{2}Pennsylvania State University \\
\texttt{\{zengyif, tendolko, baartmar, lizhong.chen, huazheng.wang\}@oregonstate.edu} \\
\texttt{\{qingyun.wu\}@psu.edu} \\
}

\begin{document}
\maketitle
\def\thefootnote{*}\footnotetext{Equal Contribution.}\def\thefootnote{\arabic{footnote}}

\begin{abstract}

Ranking passages by prompting a large language model (LLM) can achieve promising performance in modern information retrieval (IR) systems.
A common approach to sort the ranking list is by prompting LLMs for a pairwise or setwise comparison which often relies on sorting algorithms. 
However, sorting-based methods require consistent comparisons to correctly sort the passages, which we show that LLMs often violate.
We identify two kinds of intrinsic inconsistency in LLM-based pairwise comparisons: \textit{order inconsistency} which leads to conflicting results when switching the passage order, and \textit{transitive inconsistency} which leads to non-transitive triads among all preference pairs.
Our study of these inconsistencies is relevant for understanding and improving the stability of any ranking scheme based on relative preferences.
In this paper, we propose \rankfusion, an LLM-based ranking framework that mitigates these inconsistencies and produces a robust ranking list. 
\rankfusion mitigates order inconsistency using in-context learning (ICL) to demonstrate order-agnostic comparisons and calibration to estimate the underlying preference probability between two passages.
We then address transitive inconsistency by aggregating the ranking results from multiple rankers.
In our experiments, we empirically show that \rankfusion can significantly reduce inconsistent comparison results, improving the ranking quality by making the final ranking list more robust.
Our code is available at \href{https://github.com/XHMY/LLM-RankFusion}{https://github.com/XHMY/LLM-RankFusion}

\end{abstract}

\section{Introduction}

\begin{figure*}[ht]
    \centering
    \vspace{-3mm}
    \includegraphics[width=\textwidth]{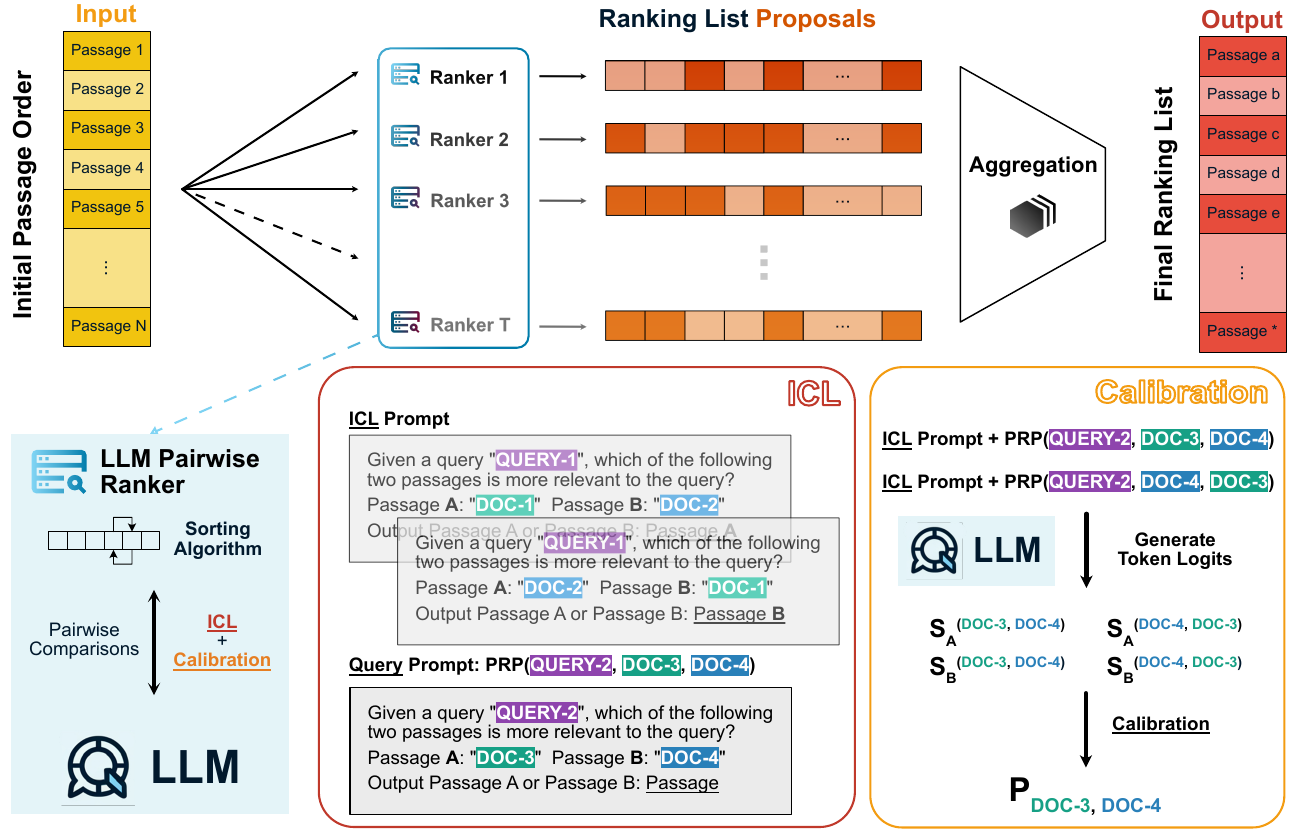}
    \caption{The \rankfusion pipeline.
    It shows an illustration of the aggregation process to mitigate the transitive inconsistency.
    The ranking list proposals are formed by different rankers, and the details of each ranker are shown in the lower left of the figure.
    Each ranker includes ICL and calibration to address order inconsistency.
    }
    \label{fig:rankfusion}
    \vspace{-3mm}
\end{figure*}

Large language models (LLMs) have demonstrated strong zero-shot and few-shot capabilities in many natural language processing tasks \cite{achiam2023gpt,zeng2024autodefense,zhao2023survey}.
This enables the effective integration of these LLMs in modern information retrieval (IR) systems \cite{wu2023survey,li2024survey}.
Without supervised training on labeled data in a specific task, LLMs can adapt to the task by prompt and pipeline design.
Recent work has tried to apply LLMs in text ranking and shown promising performance \cite{qin2023large,chao2024make,sun2023rankgpt,ma2023zero}.
Text ranking is an important task in modern recommendation systems and search engines, which refines the order of the retrieved passages to improve the output quality \cite{liu2009learning}.
Traditional supervised text ranking methods mostly rely on large amounts of human-annotated labels \cite{bajaj2016ms,bonifacio2021mmarco,thakur2021beir,nogueira2019passage}.
Typical LLM-based ranking approaches prompt LLMs to generate partial orders, comparisons, or relevance scores without additional training in advance.

Despite the great potential of LLMs in passage ranking, they can also suffer from significant inconsistency.
Previous work shows LLM-based ranking is sensitive to the order of input passages in the prompt, which stems from the positional bias of LLMs \cite{wang2023large,lu2021fantastically,tang2023found}.
In a pairwise comparison between two passages, the results can conflict before and after swapping the passages.
We identify this as \textbf{order inconsistency}.
Even if we can fully mitigate the order inconsistency within a single comparison, there is still a concern about inconsistency among multiple different comparisons.
For example, PRP-Sort \cite{qin2023large} uses sorting algorithms to efficiently produce a full ranked list from pairwise comparisons.
However, these sorting algorithms typically assume the transitivity of comparisons to correctly sort the passage, which we show that LLMs often cannot guarantee.
(e.g. $d_1 \succ d_2, d_2 \succ d_3 \Rightarrow d_1 \succ d_3$, where $d_i$ represents a passage and $\succ$ means "preferred to").
We identify this as \textbf{transitive inconsistency}, which has been ignored in previous works.
Due to the transitive inconsistency, we show that the ranker's performance is highly sensitive to the initial input order of the retrieved documents.
Ranking systems expect to produce a robust ranking list to present to users, but if different initial orderings can lead to significant variant ranked lists, the reliability of the LLM's rankings can be brought into question.
Our analysis of this inconsistency can be easily extended to setwise and listwise comparisons, as both ranking schemes rely on relative preferences and are fundamentally pairwise in nature. 
This broad applicability highlights that identifying and addressing these inconsistencies in pairwise ranking is relevant for improving the reliability of any ranking scheme based on relative preferences.
We propose the \rankfusion framework as shown in Figure \ref{fig:rankfusion} to produce a consistent ranking list by mitigating the above inconsistencies.
To mitigate order inconsistency, we first use calibration to address the conflict before and after swapping the passages.
It calculates the preference probability based on the logit output, which then produces the preference without bias in position.
To further let LLM realize the preference should be order agnostic, we propose the in-context learning (ICL) ranking prompt.
The ICL prompt uses an example to demonstrate the swapping of passages doesn't affect the preference.
While we can greatly reduce order inconsistency by ICL and calibration, the improved pairwise comparison still does not address transitive inconsistency directly.
We use rank aggregation to further mitigate the transitive inconsistency.
Rank aggregation is a commonly used post-process method in combining multiple ranking lists and yielding a robust ranking result \cite{dwork2001rank, lin2010rank, schalekamp2009rank}.
We aggregate the ranking list proposals from LLM-based pairwise and setwise rankers with different underlying sorting algorithms or different LLMs that are responsible to make preference decisions.

In experiments, we show that ICL and calibration can reduce the effect of positional bias and increase the NDCG score significantly.
We then show the effectiveness of aggregation in addressing transitive inconsistency by forming a consensus from multiple ranking lists.
In summary, the contributions of this paper are:
\begin{itemize}
    \item We investigate order inconsistency and, for the first time, identify and study transitive inconsistency in LLM-based ranking.
    \item We address order inconsistency by ICL and calibration. The improvement is significant in most LLMs.
    \item We bridge the area of LLM-based ranking with rank aggregation to mitigate the impact of transitive inconsistency in pairwise and setwise comparisons.
    \item We show the promising empirical performance of the aggregation method by studying the aggregation among different sorting algorithms and LLMs.
\end{itemize}

\section{Related Work}

\begin{table*}[t]
    \centering
\begin{tabular}{l|rrr|rrr}
\toprule
& \multicolumn{3}{c|}{w/o ICL} & \multicolumn{3}{c}{w/ ICL} \\
& logits \texttt{A} & logits \texttt{B} & Discrepancy & logits \texttt{A} & logits \texttt{B} & Discrepancy \\
\midrule
Flan-T5-XXL & -1.37 & -0.97 & 0.20 & -1.12 & -0.86 & 0.13 \\
Flan-UL2 & -1.15 & -1.64 & -0.24 & -1.18 & -1.37 & -0.09 \\
Llama-3-8B & 6.44 & 6.56 & 0.06 & 4.46 & 3.93 & -0.26 \\
Llama-3-70B & -0.15 & 2.99 & 0.92 & 0.31 & 0.17 & -0.07 \\
Gemma-7B & 280.75 & 273.24 & -1.00 & 292.60 & 291.85 & -0.36 \\
Vicuna-13B & 16.75 & 15.88 & -0.41 & 20.10 & 20.05 & -0.03 \\
Mixtral-8x7B & 26.85 & 19.93 & -1.00 & 28.01 & 26.19 & -0.72 \\
GPT-3.5-Turbo & -4.54 & -5.23 & -0.33 & -6.25 & -5.57 & 0.33 \\
GPT-4 & -4.71 & -4.54 & 0.09 & -3.74 & -5.97 & -0.81 \\
\bottomrule
\end{tabular}
    \caption{
    Average logits value of the token \texttt{A} and the token \texttt{B} from pairwise comparisons.
    The token with the highest logits will be sampled as the generated output.
    The Discrepancy is the discrepancy between the $softmax$ of these two logits values.
    The absolute value of the discrepancy close to 1 means the LLM is very biased on position.
    The ICL stands for in-context learning, which is further elaborated in Section \ref{sec:icl}. 
    }
    \label{tab:logprob_oder_inconsistency}
\end{table*}
\begin{table*}[t]
    \centering
    \begin{tabular}{lrrrr}
\toprule
Model & \# Circular Triads & \# Type-1 Triads & \# Type-2 Triads & \# Total Inconsistent Triads \\
\midrule
Flan-T5-XXL & 104.67 & 4720.67 & 708.72 & 5534.07 \\
Flan-UL2 & 50.53 & 4875.74 & 496.53 & 5422.81 \\
Llama-3-8B & 28.70 & 5556.67 & 623.44 & 6208.81 \\
Llama-3-70B & 10.02 & 4533.98 & 199.09 & 4743.09 \\
Gemma-7B & 0.72 & 13326.53 & 155.65 & 13482.91 \\
Vicuna-13B & 16.40 & 8475.70 & 460.63 & 8952.72 \\
Mixtral-8x7B & 20.70 & 8970.58 & 672.49 & 9663.77 \\
GPT-3.5-Turbo & 40.30 & 6849.77 & 705.56 & 7595.63 \\
GPT-4 & 78.65 & 3694.65 & 754.91 & 4528.21 \\


\bottomrule
\end{tabular}
    \caption{The number of inconsistent triads in the tournament graph.
    A higher number indicates more inconsistency.
    Type-1 triads refer to $d_i = d_j, d_j = d_k, d_k \succ d_i$.
    Type-2 triads refer to $d_i = d_j, d_i \succ d_k, d_k \succ d_j$.
    The total inconsistent triads refer to the sum of all kinds of inconsistent triads.
    The maximum number of inconsistent triads for 100 passages is 161684 \cite{kulakowski2018inconsistency}.
    This tournament graph is constructed from the comparisons without ICL and calibration.
    }
    \label{tab:traid-logical-inconsistency}
\end{table*}
\noindent\textbf{LLM-based ranking approaches} have been developed with distinct ranking schemes.
Pointwise approaches \cite{liang2023pointwiseyesno, sachan2023pointwiseqlm, drozdov2023parade} aim to estimate the relevance between a query and a single document.
Listwise \cite{sun2023rankgpt,ma2023zero} ranking methods aim to directly rank a list of documents by inserting the query and document list into an LLM's prompt and instructing it to output the reranked document identifiers, though they rely on the strong capability of LLMs, suffer from positional bias and are sensitive to document order in the prompt \cite{zhu2023large}.
Pairwise ranking methods \cite{qin2023large} provide the query and a pair of documents to the LLM, which is instructed to generate the identifier of the more relevant document; these pairwise comparisons are then aggregated using efficient sorting algorithms like Heapsort or Bubblesort to produce the final ranking.
The Setwise approach \cite{zhuang2023setwise} is also proposed to compare a set of documents at a time to further improve efficiency.

\noindent\textbf{Rank aggregation} has been widely used in many information retrieval tasks \cite{farah2007outranking,dwork2001rank}.
Previous works \cite{pradeep2021expando,gienapp2022sparse} also use aggregation to form the ranking list from pairwise comparison.
We employ Borda count \cite{borda1784} to aggregate different ranking lists in our paper.
Borda count assigns a score to each item based on its position in each input ranking and sums these scores to produce the final aggregated ranking.
While other rank aggregation methods exist, such as Markov Chain-based methods \cite{dwork2001rank}, supervised learning approaches \cite{liu2007supervised}, and unsupervised methods like Kemeny rank aggregation \cite{kemeny1962preference}, many of these are either NP-hard or not specifically designed for list aggregation.
By using the simple and computationally efficient Borda count method, we aim to demonstrate the power of combining LLM-based ranking with aggregation, even without resorting to more complex aggregation techniques.

\section{Inconsistency  of LLM-based ranking} \label{sec:preliminary}

\subsection{Inconsistency of Pairwise Comparisons} \label{inconsistency}

In this work, we identify and distinguish two types of inconsistencies that LLM-based rankers exhibit:

\begin{enumerate}
    \item \textbf{Order Inconsistency:} The LLM’s judgment on a pair of passages changes depending on the order they are presented in the prompt, which is also known as positional bias \cite{lu2021fantastically}. 
    \item \textbf{Transitive Inconsistency:} The LLM makes a series of three or more judgments that logically contradict each other, over a set of three or more passages, i.e., $d_1 \succ d_2, d_2 \succ d_3, d_3 \succ d_1$.
\end{enumerate}

Under the pairwise ranking approach, each LLM query produces a pairwise comparison results on $d_i, d_j$, the result can be $d_i \succ d_j$ or $d_j \succ d_i$, where $d$ represents a passage and $\succ$ means "preferred to".
While we focus on pairwise comparisons in this work, our analysis of these two types of inconsistency can be easily extended to setwise or listwise ranking schemes.
This is because both setwise and listwise comparisons still rely on relative preferences between individual passages, and are therefore fundamentally pairwise in nature.
For example, the result of a setwise comparison $d_i \succ d_j, d_k$ can be represented as two implicit pairwise comparisons, $d_i \succ d_j$ and $d_i \succ d_k$.
Therefore, our inconsistency analysis can not only be applied to setwise and listwise ranking, but any ranking scheme that involves relative comparisons between passages.

\subsection{Inconsistency Measurement}

We measure the inconsistency in comparisons of a variety of LLMs using the TREC-DL2019 test set.
We construct the pairwise preference among all passages using PRP-Allpair \cite{qin2023large}.
The order inconsistency can be shown in Table \ref{tab:logprob_oder_inconsistency}.
In a pairwise ranking scheme, we ask the LLM to output \texttt{A} to select the first passage or \texttt{B} to select the second passage.
We query both the permutations of passages and collect the logits of token \texttt{A} and token \texttt{B}.
For an LLM with no positional bias, switching the order of a pair of passages will not affect the preference judgment, which leads to the average logits of \texttt{A} and \texttt{B} being equal.
However, we can observe from Table \ref{tab:logprob_oder_inconsistency} that the logits of these tokens typically suffer from an obvious discrepancy.
This indicates that the LLM's choice is often biased towards either \texttt{A} or \texttt{B}, which implies order inconsistency.


We represent pairwise comparisons using a tournament graph - a complete graph where each vertex represents a passage and edges represent preferences between passages. In this graph, directed edges $d_i \rightarrow d_j$ indicate that passage $d_i$ is preferred over $d_j$, while undirected edges indicate ties. Following PRP-Sort \cite{qin2023large}, we handle cases of order inconsistency (where swapping passage positions leads to different preferences) by marking them as ties in the graph. While these ties are treated equally in the graph structure, we note that the underlying passages may have different ground truth relevance scores.

To measure transitive inconsistency, we first construct the tournament graph by performing all pairwise comparisons between passages and marking order inconsistent pairs (those that change preference when swapped) as ties. We then count the number of inconsistent triads in the graph, following the method of \citet{kulakowski2018inconsistency}. An inconsistent triad occurs when three passages form a cycle of strict preferences - for example, when $d_1$ is preferred over $d_2$, $d_2$ is preferred over $d_3$, but $d_3$ is preferred over $d_1$, violating transitivity. The count of such triads serves as our metric for transitive inconsistency. As shown in Table \ref{tab:traid-logical-inconsistency}, we observe that the frequency of inconsistent triads varies across different LLMs, with larger models generally exhibiting fewer transitive inconsistencies.

\subsection{The Impact of Inconsistency}

\begin{table}[t]
\centering
\begin{tabular}{lcccc}  \toprule
& \multicolumn{2}{c}{Bubblesort} & \multicolumn{2}{c}{Heapsort}\\
& BM25 & Hard List & BM25 & Hard List \\ \midrule
Flan-T5-XXL & 67.87 & 27.54 & 70.65 & 67.42 \\ 
Flan-UL2 & 72.63 & 42.33 & 72.45 & 69.95 \\ 
Llama-3-8b & 65.38 & 12.61 & 68.46 & 65.41 \\ 
Llama-3-70b & 72.43 & 35.97 & 73.71 & 71.00 \\ 
Gemma-7b & 61.25 & 15.00 & 50.70 & 49.47 \\ 
Vicuna-13b & 50.59 & 6.15 & 61.43 & 59.05 \\ 
Mixtral-8x7b & 65.06 & 13.65 & 69.14 & 66.58 \\ 
GPT-3.5-Turbo & 64.72 & 12.52 & 67.69 & 61.54  \\ 
GPT-4 & 72.04 & 39.56 & 73.48 & 72.71 \\

\bottomrule
\end{tabular}
\caption{
NDCG@10 of ranking from different initial orders.
Hard list is the inverse order of the ranking result obtained by performing the PRP-Allpair ranking scheme \cite{qin2023large}.
Note that Heapsort results are largely unaffected, suggesting that this specific hard list is for Bubblesort only. 
}
\label{tab:hard-list}
\end{table}

As shown in Tables \ref{tab:logprob_oder_inconsistency} and \ref{tab:traid-logical-inconsistency}, LLM-based rankers can exhibit significant amounts of inconsistency across judgments.
Applying sorting-based ranking schemes based on non-transitive pairwise comparisons can produce volatile result rankings that are highly sensitive to the initial order of candidate passages.
This can have particularly adverse effects if that initial ordering is a "hard list" for the chosen sorting method, as demonstrated in Table \ref{tab:hard-list}.
A hard list is an initial order of passages where the high-relevance passages require many comparisons to be moved to the front of the ranking, increasing the likelihood of encountering a transitive inconsistency which blocks the promotion of the passage.
A hard list for one sorting algorithm may not be as hard of a list for another, which is demonstrated in Table \ref{tab:hard-list}.
By aggregating the full ranked lists from multiple sorting algorithms, we can mitigate the worst-case effects of inconsistency while producing more robust final rankings.

\section{Addressing LLM-based Ranking Inconsistency} \label{sec:methodology}

\subsection{Mitigating Order Inconsistency} 

LLMs suffer from positional bias, which leads to the order inconsistency.
This will result in conflicting comparisons after swapping the passage position.
Previous work handled order inconsistency as ties in the comparison, which ignores the positional bias nature of LLM-based ranking.
We propose 2 methods to mitigate the order inconsistency in the \rankfusion.

\subsubsection{In-Context Learning (ICL)} \label{sec:icl}

We design the ICL prompt to utilize the few-shot ability \cite{brown2020language} of LLMs to mitigate order inconsistency.
The prompt provides the LLM with an example pairwise comparison for both order permutations as shown in Figure \ref{fig:rankfusion}.
This demonstration illustrates that the task is to compare the passages based on their relevance to the query instead of its position in the prompt.
As shown in Table \ref{tab:logprob_oder_inconsistency}, using ICL can balance the probability of LLM selecting a passage from either position.

\subsubsection{Calibration}

In a pairwise ranking scheme, we ask the LLM to output token \texttt{A} to select the first passage or \texttt{B} to select the second passage in the given prompt.
The positional bias makes LLMs more likely to select the passage in a certain position of the prompt instead of only based on relevance.
By considering the comparisons from both possible positions, even if the LLM is biased to a specific position, using the average output probability of a passage under all positions may address this bias.

For every pair of passages, we query the LLM with two permutations $(A =d_i, B= d_j)$ and $(A =d_j, B= d_i)$.
For each output, \texttt{A} represents the first passage and \texttt{B} represents the second passage in the prompt.
We then take the token probability of \texttt{A} and \texttt{B} from the LLM.
We denote $S_A^{(ij)}$ and $S_B^{(ij)}$ as the log probabilities of the output tokens \texttt{A} and \texttt{B} with the permutation  $(A =d_i, B= d_j)$ . 
$P_{ij}^{(ij)} \in [0, 1]$ is the probability of $d_i$ preferred to $d_j$ with the permutation  $(A =d_i, B= d_j)$ .
The probability is generated among all candidate tokens by LLM.
We know that only token \texttt{A} and token \texttt{B} represent a valid choice of passage, so we compute $P_{ij}^{(ij)}$ and  $P_{ji}^{(ji)}$ as following:
\begin{equation}
P_{ij}^{(ij)} = \frac{e^{S_A^{(ij)}}}{e^{S_A^{(ij)}} + e^{S_B^{(ij)}}} ,\quad
P_{ji}^{(ji)} = \frac{e^{S_A^{(ji)}}}{e^{S_A^{(ji)}} + e^{S_B^{(ji)}}}
\end{equation}
To make the comparison satisfy $P_{ij} + P_{ji} = 1$, we denote the calibrated  probability of $d_i \succ d_j$ as
\begin{equation}
P_{ij} = \frac{e^{P_{ij}^{(ij)}}}{e^{P_{ij}^{(ij)}} + e^{P_{ji}^{(ji)}}}
\end{equation}
Finally, we obtain the calibrated preference of $d_i \succ d_j$ if $P_{ij}>0.5$.

\subsection{LLM Ranking Aggregation}

In passage ranking, we can generate multiple proposed ranking lists using different ranking settings.
These settings can include different sorting algorithms to build fully ranked lists from pairwise, setwise, or listwise comparisons, as well as other factors like a specific LLM used for the preference query.
Given the inconsistency discussed in \ref{sec:preliminary}, we know that any configuration of these settings can result in noisy rank results.
We cannot assume we know \textit{a priori} which proposed rank setting is the best, so choosing a single setting that will be affected the least by inconsistency is difficult.

To address noisy ranked list proposals, we propose \rankfusion, a rank aggregation pipeline as shown in Algorithm \ref{alg:agg-pipeline}.
Rank aggregation can address conflicting results by combining these results into a single, coherent ranking, limiting the effects of noisy settings.

Let $C = \{d_1, d_2, \ldots, d_m\}$ be the set of $m$ passages, and let $V = \{v_1, v_2, \ldots, v_n\}$ be the set of $n$ ranked list proposals.
Each proposal $v_j$ ranks the passages in order of preference, producing a ranked list $L_j$.
We apply Borda count \cite{shah2018simple,emerson2013original,borda1784} to aggregate these proposals. For each passage $d_i$ is calculated as follows:

\begin{equation}
B(d_i) = \sum_{j=1}^{n} (m - r_{ij})
\end{equation}

where $r_{ij}$ is the rank of passage $d_i$ in voter $v_j$'s list $L_j$. The passage with the highest Borda count is the winner.
The aggregated list $L$ is obtained by sorting the passages in descending order of their Borda counts:

\begin{equation}
L = \langle d_{\sigma(1)}, d_{\sigma(2)}, \ldots, d_{\sigma(m)} \rangle
\end{equation}

where $\sigma$ is a permutation of $\{1, 2, \ldots, m\}$ such that:

\begin{equation}
B(d_{\sigma(1)}) \geq B(d_{\sigma(2)}) \geq \ldots \geq B(d_{\sigma(m)})
\end{equation}

\begin{algorithm}[t]
\caption{LLM Rank Aggregation Pipeline} \label{alg:agg-pipeline}
\begin{algorithmic}[1]
\Require{Query $q$, Corpus $D$, Rank settings ${R_1, R_2, \ldots, R_k}$}
\Ensure{Aggregated rank list $L$}
\State $L_1, L_2, \ldots, L_k \gets \emptyset$ \Comment{Initialize empty rank lists}
\For{$i \gets 1$ to $k$}
\State $L_i \gets$ \Call{Ranking}{$q$, $D$, $R_i$} \Comment{Generate rank list using rank setting $R_i$}
\EndFor
\State $L \gets$ \Call{RankAggregation}{$L_1, L_2, \ldots, L_k$} \Comment{Aggregate rank lists}
\State \Return $L$
\end{algorithmic}
\end{algorithm}

\subsubsection{Aggregation Across Sorting Methods}

We aggregate the ranked list proposals from two comparison-based sorting algorithms, Bubblesort and Heapsort.
Bubblesort repeatedly steps through the list, compares adjacent elements, and swaps them if they are in the wrong order;
Heapsort uses a binary heap to sort elements.
By combining the ranked lists from algorithms with different properties, the aggregated result becomes more robust to variations in the input data.
If one algorithm performs poorly on a particular input, the other algorithm may compensate for it, leading to a more consistent overall ranking.

\subsubsection{Aggregation Across LLMs}

Individual LLMs might also have unique biases in their preferences and, therefore, unique transitive inconsistency.
This motivates aggregation across ranking lists from multiple LLMs.
This can help to reduce the impact of any individual LLM, which may be inconsistent in handling certain queries.
The aggregated result formed by decisions from multiple LLMs can be more robust and consistent.
We group LLMs by their size and aggregate the ranking lists from LLMs with the similar number of parameters.




\begin{table*}[ht]
\centering
\begin{tabular}{lrrrr}
\toprule
Model & Baseline & ICL Only & Calibration Only & ICL + Calibration \\
\midrule
Flan-T5-XXL & 67.87 & 68.64 ({\color{teal}+0.77}) & 69.73 ({\color{teal}+1.86}) & 71.05 ({\color{teal}+3.18}) \\
Flan-UL2 & 72.63 & 70.14 ({\color{red}-2.49}) & 72.53 ({\color{red}-0.10}) & 73.70 ({\color{teal}+1.07}) \\
Llama-3-8B & 65.38 & 66.35 ({\color{teal}+0.97}) & 69.58 ({\color{teal}+4.20}) & 71.51 ({\color{teal}+6.13}) \\
Llama-3-70B & 72.43 & 71.62 ({\color{red}-0.81}) & 74.12 ({\color{teal}+1.69}) & 74.55 ({\color{teal}+2.12}) \\
Gemma-7B & 56.35 & 61.04 ({\color{teal}+4.69}) & 60.18 ({\color{teal}+3.83}) & 63.11 ({\color{teal}+6.76}) \\
Vicuna-13B & 61.25 & 68.07 ({\color{teal}+6.82}) & 64.46 ({\color{teal}+3.21}) & 70.42 ({\color{teal}+9.17}) \\
Mixtral-8x7B & 65.06 & 70.05 ({\color{teal}+4.99}) & 66.75 ({\color{teal}+1.69}) & 71.16 ({\color{teal}+6.10}) \\
GPT-3.5-Turbo & 64.72 & 68.83 ({\color{teal}+4.11}) & 68.99 ({\color{teal}+4.27}) & 72.06 ({\color{teal}+7.34}) \\
GPT-4 & 72.04 & 73.34 ({\color{teal}+1.30}) & 74.56 ({\color{teal}+2.52}) & 74.79 ({\color{teal}+2.75}) \\


\bottomrule
\end{tabular}
\caption{
NDCG@10 of PRP-Sorting with Bubblesort, starting from the BM25 initial order on TREC DL 2019. 
The experiment shows an ablation study of using ICL and calibration to improve ranking performance by addressing order inconsistency.
}
\label{tab:exp-icl-cal}
\end{table*}

\begin{table}[]
\centering
\begin{tabular}{lll}
\toprule
& DL19 & DL20 \\
\midrule
Setwise & 63.25 & 60.27 \\
Listwise (RankGPT) & 69.61 & 65.49 \\
Pairwise (PRP-Sort) & 65.38  & 60.16 \\
Pairwise+ICL+Calibration (Ours) & 71.51  & 65.10 \\
\bottomrule
\end{tabular}
\caption{NDCG@10 on TREC DL 2019 and 2020 datasets using Llama-3-8B model. Our proposed method (Pairwise+ICL+Calibration) shows superior performance compared to setwise, listwise, and standard pairwise approaches.}
\label{tab:exp-baselines-comp}
\end{table}

\section{Experiments}

\subsection{Experimental Setup} \label{sec:experimental_setup}

We utilize test sets from TREC, a standard dataset for information retrieval research.
Specifically, we use the top 100 passages retrieved by BM25 \cite{lin2021pyserini,robertson2009probabilistic} for each of the queries associated with the TREC-DL2019 and 2020 test sets \cite{craswell2020overview}.
Our results are based on the re-ranking of these 100 passages.
The LLM ranking scheme is implemented under the same experimental setting of PRP-Sort \cite{qin2023large}.

We evaluate our results using two metrics: Normalized Discounted Cumulative Gain (NDCG) and average Kendall tau Distance. 
NDCG is a standard metric used to evaluate the quality of ranked retrieval results.
It accounts for the position of relevant documents in the ranking, assigning higher importance to documents appearing earlier. 
Kendall tau distance is a metric used to measure the dissimilarity between two rankings.
We compare our results against pairwise, setwise, and listwise baselines \cite{qin2023large, zhuang2023setwise, sun2023rankgpt}.


\subsection{Addressing Order Inconsistency}

We have shown that in-context learning can help to balance the average probability of the two choices given during pairwise ranking in Table \ref{tab:logprob_oder_inconsistency}.
In Table \ref{tab:exp-icl-cal}, we can see that solely using ICL can help improve the ranking performance in most LLMs.
Some encoder-decoder models like Flan-UL2 may intrinsically suffer from less positional bias than other decoder-only LLMs \cite{liu2024lost}, so we don't expect much improvement with using ICL for those models.
The calibration addresses order inconsistency by calculating preference probability based on comparison from both positions.
The improvement from solely using the calibration is also significant, as shown in Table \ref{tab:exp-icl-cal}.
We also find that using these two methods individually cannot improve ranking performance for Flan-UL2, but combining them can bring improvement compared to the baseline.
Furthermore, as demonstrated in Table \ref{tab:exp-baselines-comp}, our approach of combining pairwise ranking with ICL and calibration outperforms several baseline methods across different datasets.

\begin{table*}[]
\centering
\begin{tabular}{llrrr}
\toprule
&  & Flan-T5-XXL & Llama-3-8B & \textbf{Model Aggregation} \\
\midrule
\multirow{5}{*}{TREC DL 2019} & Pairwise.BubbleSort & 71.05 & 71.51 &  73.08\\
& Pairwise.HeapSort & 69.76 & 70.82 &   72.24\\
& Setwise.BubbleSort & 71.07 & 63.25 &   64.01\\
& Setwise.HeapSort & 70.46 & 65.14 &   67.32\\
& \textbf{Scheme Aggregation} &  70.78&  71.61&   \\
\midrule
\multirow{5}{*}{TREC DL 2020} & Pairwise.BubbleSort & 69.96 & 65.10 &   70.95\\
& Pairwise.HeapSort & 69.61 & 64.01 &   69.88\\
& Setwise.BubbleSort & 68.66 & 60.27 &   61.73\\
& Setwise.HeapSort & 68.82 & 60.68 &   63.9\\
& \textbf{Scheme Aggregation} &  69.74&  65.09&  \\
\bottomrule
\end{tabular}
\caption{NDCG@10 of different ranking schemes and model aggregations on TREC DL 2019 and 2020 datasets. The table shows individual performance of Flan-T5-XXL and Llama-3-8B models, as well as their aggregation, across various ranking schemes.
The pairwise ranking results in this table both apply ICL and calibration.
The initial ranking list is from the BM25 order.
}
\label{tab:scheme-model-agg}
\end{table*}

\subsection{Addressing Inconsistency via Aggregation}

Table \ref{tab:scheme-model-agg} demonstrates the effectiveness of both model and ranking scheme aggregation in improving ranking performance. Model Aggregation consistently outperforms individual models across both datasets, particularly for pairwise ranking.
This trend is consistent across different ranking schemes and datasets, showing robustness through aggregation.
On ranking scheme aggregation, we observe that combining different approaches mitigates individual weaknesses.
Scheme Aggregation shows balanced performance, often achieving scores close to or exceeding the best individual scheme.
This suggests that aggregating across ranking schemes can provide a more stable and potentially superior ranking, leveraging the strengths of both pairwise and setwise approaches, as well as different sorting algorithms.

\begin{table*}[]
\centering
\begin{tabular}{lcccc|cc} \toprule
& \multicolumn{2}{c}{w/o ICL + Calibration} & \multicolumn{2}{c}{w/ ICL + Calibration} & \multicolumn{1}{c}{{\rankfusion}}\\
 & Bubblesort & Heapsort & Bubblesort & Heapsort & Aggregated \\
\midrule

Flan-T5-XXL & 62.93 $\pm$ 1.48 & 68.56 $\pm$ 0.77 &  70.05 $\pm$ 0.65 & 70.32 $\pm$ 0.38 & 70.45 $\pm$ 0.47\\
Flan-UL2  &  68.80 $\pm$ 1.06 & 71.68 $\pm$ 0.53 & 72.62 $\pm$ 0.45 & 72.95 $\pm$ 0.33 & 72.88 $\pm$ 0.34\\
Llama-3-8B &  53.72 $\pm$ 2.34 & 65.73 $\pm$ 0.97 & 70.16 $\pm$ 0.53 &  70.57 $\pm$ 0.35 & 70.58 $\pm$ 0.45\\
Llama-3-70B &  68.06 $\pm$ 1.18 & 72.69 $\pm$ 0.63 & 73.30 $\pm$ 0.48 &  73.64 $\pm$ 0.20 & 73.59 $\pm$ 0.28\\
Gemma-7B &  30.94  $\pm$ 3.52 & 45.26 $\pm$ 1.85 &   58.56 $\pm$ 0.96 & 59.84 $\pm$ 0.59 & 59.63 $\pm$ 0.62\\
Vicuna-13B & 45.88 $\pm$ 2.70 & 58.99 $\pm$ 0.96 & 68.19 $\pm$ 0.59 &  69.02 $\pm$ 0.45 & 68.90 $\pm$ 0.49\\
Mixtral-8x7B & 52.58 $\pm$ 2.52 & 67.27 $\pm$ 1.01 & 67.36 $\pm$ 0.84  & 70.88 $\pm$ 0.62 & 69.76 $\pm$ 0.69\\
GPT-3.5-Turbo &  51.83 $\pm$ 2.70 & 64.68 $\pm$ 1.10 &  70.63 $\pm$ 0.62 & 71.18 $\pm$ 0.49 & 71.09 $\pm$ 0.54\\
GPT-4 &  69.93 $\pm$ 0.96 & 73.45 $\pm$ 0.48 &  74.22 $\pm$ 0.50 & 74.75 $\pm$ 0.16 & 75.04 $\pm$ 0.24\\
\bottomrule
\end{tabular}
\caption{
NDCG@10 of aggregating across sorting algorithms with standard deviation among 100 experiments on TREC DL 2019.
}
\label{tab:exp-ablation-ndcg}
\end{table*}

\begin{table*}[]
    \centering
\begin{tabular}{lcccc} \toprule
& \multicolumn{2}{c}{Bubblesort} & \multicolumn{2}{c}{Heapsort}\\
& KT & NDCG@10 & KT & NDCG@10 \\
\midrule
Flan-T5-XXL & 0.190 & 70.05 $\pm$ 0.65 & 0.075 & 70.32 $\pm$ 0.38\\
Llama-3-8B &  0.130 & 70.16$\pm$ 0.53 & 0.060 & 70.57$\pm$ 0.35\\
Gemma-7B &  0.187& 58.56 $\pm$ 0.96 & 0.075   & 59.84$\pm$ 0.59\\
Small-Model \rankfusion & 0.187  & 69.50 $\pm$ 0.56 & 0.062& 71.32 $\pm$ 0.35\\
\midrule
Flan-UL2  &  0.165 & 72.62 $\pm$ 0.45  & 0.069  & 72.95 $\pm$ 0.33\\
Vicuna-13B &  0.151 & 68.19 $\pm$ 0.59 & 0.067 & 69.02$\pm$ 0.45\\
Mixtral-8x7B &  0.299 & 67.36 $\pm$ 0.84 & 0.123  & 70.88$\pm$ 0.62\\
Medium-Model \rankfusion &0.226 & 70.93 $\pm$ 0.64 &0.076 & 73.09$\pm$ 0.42\\
\midrule
Llama-3-70B &  0.147 & 73.30$\pm$ 0.48 & 0.066& 73.64 $\pm$ 0.20\\
GPT-3.5-Turbo &  0.172 & 70.63$\pm$ 0.62 & 0.076 & 71.18$\pm$ 0.49\\
GPT-4 &  0.229 & 74.82 $\pm$ 0.31 & 0.142 & 75.02$\pm$ 0.16\\
Large-Model \rankfusion & 0.197 & 74.22$\pm$ 0.50 & 0.084 & 74.75$\pm$ 0.26\\
\bottomrule
\end{tabular}
\caption{ Aggregation results across 100 initial orderings for multiple models based on respective size on TREC DL 2019. It shows that the multi-model aggregation also balances out worse NDCG@10 and Kendall-tau distances associated with specific models, and provides a more comprehensive ranking result taking multiple factors into account. The bubblesort and heapsort data associated with each model refer to after ICL and calibration have been applied.
}
\label{tab:llm_aggregation}
\end{table*}

\paragraph{Aggregation across sorting algorithms} attempts to mitigate the impact of the initial order on an individual sorting algorithm.
As seen in Table \ref{tab:exp-ablation-ndcg} and Table \ref{tab:exp-ablation-kt} in the appendix,  \rankfusion provides significant benefits to baseline sorting-based ranking approaches.
Each of the initial orderings underwent re-ranking from both pairwise prompting with Bubblesort and Heapsort, and the two final rank lists were aggregated using Borda count.
NDCG@10 and Kendall-tau distance were recorded and averaged across each final list for each model.
In all models, we see an increase in NDCG@10 as well as reduced inconsistency with lower Kendall-tau distances.
ICL and calibration significantly increase the NDCG while also reducing the Kendall-tau distance, and the aggregation further reduces the inconsistency of the worse sorting algorithm (in this case Bubblesort), while keeping NDCG relatively constant.
This is a promising result, showing that \rankfusion can improve both consistency and ranking quality by aggregating the rankings produced by different sorting algorithms.

\paragraph{Aggregation across LLMs} attempts to produce a final ranking that is less sensitive to the individual biases of different LLMs.
This can be observed by comparing Table \ref{tab:llm_aggregation} to previous Table \ref{tab:exp-ablation-ndcg} and Table \ref{tab:exp-ablation-kt} in the appendix, as the average Kendall-tau distance of the aggregated rankings can always achieve a medium performance among the LLMs it aggregates from.
This shows promising results of \rankfusion to find a consensus ranking list that achieves \emph{balanced or even superior} performance.
In a typical passage ranking case, we don't assume we know which LLM is better at ranking the specific passage list.
Hence, it is particularly useful if we can get at least medium performance rankings by aggregating multiple proposals.

\subsection{Computation Cost} \label{sec:performance}

The performance of LLM-based passage ranking is mainly determined by the LLM inference performance and the number of comparisons required to sort a list.
The aggregation in \rankfusion requires that takes in individual ranking lists and the time cost depends on the slowest model if individual rankings are computed in parallel.
The calibration doesn't bring additional inference costs because it only collects the output logits and calculates preference probability based on two permutations of passages.
The ICL leads to a longer prompt, which slightly increases the computation during inference as shown in Table \ref{tab:bench} in the appendix.

\section{Conclusion} \label{sec:conclusion}

In this paper, we focus on addressing order inconsistency and transitive inconsistency we identify in LLM-based ranking.
These inconsistencies can significantly impact the reliability of LLM-based ranking systems.
To mitigate these issues, we proposed the \rankfusion pipeline, which incorporates in-context learning (ICL) and calibration to address order inconsistency and rank aggregation to tackle transitive inconsistency.
Our experiments demonstrated that ICL and calibration effectively reduce order inconsistency, leading to improved NDCG scores.
Furthermore, we showed that aggregation mitigates transitive inconsistency by forming a consensus from multiple ranking lists.
By exploring the idea of aggregating the decisions of multiple LLMs in the specific domain of passage ranking, our work highlights the potential of combining the strengths of different LLMs.

\bibliographystyle{plainnat}
\bibliography{aaai25}

\appendix
\onecolumn

\section{Technical Appendix}

\subsection{Limitations}
The aggregation process in \rankfusion requires fully sorted individual ranked lists, which increases the computation costs.
Additionally, the use of ICL leads to longer prompts, increasing the overhead during inference.
Another limitation of our study is that the experiments were conducted on limited IR datasets (TREC-DL2019 and TREC-DL2020), and other aggregation methods were not included in the current research.

\subsection{Future Work} 
Future Work can focus on LLM-based rank aggregation approaches that decide the comparison strategy on-the-fly and directly aggregate from pairwise comparisons without relying on sorting algorithms \cite{ramamohan2016dueling,heckel2019active,gienapp2022sparse}. 
Exploring the potential of aggregating LLM-based decisions in other tasks and domains beyond passage ranking could lead to a more general understanding of the effectiveness of combining the intelligence of multiple LLMs for improved performance and consistency in a wider range of applications \cite{duetting2023mechanism}.

\begin{table*}
\centering
\begin{tabular}{lcccc|cc} \toprule
& \multicolumn{2}{c}{w/o ICL + Calibration} & \multicolumn{2}{c}{w/ ICL + Calibration} & \multicolumn{1}{c}{{\rankfusion}}\\
 & Bubblesort & Heapsort & Bubblesort & Heapsort & Aggregated \\
\midrule
Flan-T5-XXL & 0.377 & 0.202 & 0.190 & 0.075 & 0.115\\
Flan-UL2  &  0.359 & 0.211 & 0.165 & 0.069 & 0.104\\
Llama-3-8B &  0.408 & 0.197 & 0.130 &  0.060 & 0.084\\
Llama-3-70B &  0.352 & 0.228 & 0.147 & 0.066 & 0.095\\
Gemma-7B &  0.493 & 0.430 & 0.187 &  0.075 & 0.115\\
Vicuna-13B &  0.444 & 0.272 & 0.151 & 0.067 & 0.096\\
Mixtral-8x7B &  0.449 & 0.290 & 0.299 & 0.123 & 0.192\\
GPT-3.5-Turbo &  0.434 & 0.242 &  0.172 & 0.076 & 0.108\\
GPT-4 &  0.332 & 0.163 & 0.229 & 0.142 & 0.173\\
\bottomrule
\end{tabular}
\caption{
Average Kendall-tau distance results in aggregating across sorting algorithms.
We don't include the standard deviation for Kendall-tau distance in this table because it is too small and insignificant.
}
\label{tab:exp-ablation-kt}
\end{table*}

\subsection{Example Passage Ranking}

We show an example passage ranking that contain 15 passages for a query.
The details of these 15 passages are shown in Table \ref{tab:bm25-passage-details-1} and Table \ref{tab:bm25-passage-details-2}.
The ranking results from different settings are shown in Table \ref{tab:rank-passage-demo}.

\begin{table}[]
\centering
\begin{tabular}{lp{0.7\linewidth}l}
\toprule
Passage Identifier & Passage Content & Relevance \\
\midrule
\relzerop{A} & {\small Well, one of Arnold's biggest insights is what resulted in the invention of the Searzall, and it's something we got wrong in our sous vide video. Sous vide, if you recall, is the process of cooking food in a controlled-temperature water bath, using a vacuum sealer to protect your meat from the liquid. What you get from sous vide is your food cooked to exactly the temperature you want to kill bacteria and make it safe to eat, but not overcooking it.} & \relzero \\
\midrule
\relthreep{B} & {\small What kind of foods can you cook sous vide? Sous vide is traditionally seen as an alternative method for cooking meats. However, the technique is extremely versatile, meaning all manner of ingredients can be cooked such as: Pork, Lamb, Beef, Chicken, Duck, Turkey, Quail.} & \relthree \\
\midrule
\reltwop{C} & {\small What is a Sous Vide Cooker? We said it before and we’ll say it again: a sous vide machine is a better way to cook. When you use a sous vide at home, you’ll be cooking with water. It’s not at all what you’re thinking. Your steak and chicken – any food really – will be placed in BPA-free, sturdy plastic bags and cooked.} & \reltwo \\
\midrule
\relzerop{D} & {\small Sous-vide cooking involves cooking food in sealed plastic bags immersed in hot water for long periods of time. Depending on the cut, type, and thickness of the meat or the type of food in question, cooking sous-vide for several hours is not out of the ordinary. The key is managing the temperature of the water so it stays hot enough to cook the food thoroughly and evenly and long enough to kill any food-borne pathogens that may be in the bag along with the food.} & \relzero \\
\midrule
\relzerop{E} & {\small Actually vacuum is a mis-represented concept in sous vide. Fact is, you don't need vacuum sealing entirely for sous vide cooking. If you can truly control the temperature of the water bath then all you need is to provide the food you are cooking a barrier from the water bath.} & \relzero \\
\midrule
\relthreep{F} & {\small Often, however, when you prepare food sous vide you’re packaging the food in plastic. (Often but not always. Eggs come in their own wrappers—their shells—and we can sous vide foods that set as they cook, like custard, yogurt, and chicken liver pâté, in glass canning jars.)} & \relthree \\
\midrule
\relzerop{G} & {\small The term sous vide (pronounced soo–veed) is a French term, meaning under vacuum. Sous vide is a culinary technique in which vacuum-sealed food is immersed in a water bath and cooked at a very precise, consistent temperature.This cooking technique typically involves cooking food for longer periods of time at a lower temperature.he term sous vide (pronounced soo–veed) is a French term, meaning under vacuum. The term s ous vide (pronounced soo–veed) is a French term, meaning under vacuum.} & \relzero \\
\midrule
\relzerop{H} & {\small The sous vide technique has been the secret of great chefs worldwide for decades. The SousVide Supreme is an amazing new all-in-one sous vide water oven designed to bring the sous vide cooking technique into home kitchens and restaurants at an affordable price. The sous vide (pronounced soo-veed) technique involves cooking food in vacuum-sealed pouches submerged in a water bath held at a precisely-controlled temperature.} & \relzero \\
\midrule
\relzerop{I} & {\small Sous vide recipes. Sous vide recipes. From the French for ‘under vacuum’, sous vide is a method of cooking where ingredients are sealed in an airtight bag and submerged in a water bath. This method not only ensures a constant cooking temperature, but allows the food to cook for long periods of time without losing any of its flavour or moisture.} & \relzero \\
\bottomrule
\end{tabular}
\caption{
Top 15 relevant passages for query \texttt{what types of food can you cook sous vide} from BM25 (Part 1 of 2).
These passages are ranked by BM25 and assigned the Passage Identifier from A to O in alphabetical order.
We refer to this as the BM25 order, which is used as the initial order for \rankfusion.
The Relevance column shows the ground truth pointwise relevance score label.
The relevance scores are from 0 to 3, where 0 is the most not relevance, 3 is the most relevance.
}
\label{tab:bm25-passage-details-1}
\end{table}

\begin{table}[]
\centering
\begin{tabular}{lp{0.7\linewidth}l}
\toprule
Passage Identifier & Passage Content & Relevance \\
\midrule
\relzerop{J} & {\small Slow and low is the name of the game here. Some times cooking food sous vide means hours, sometimes it means entire days. Food safety is a concern, but doesn't generally an issue as long as you stick to the temperature and cooking time specified for the food you're cooking.imply put, sous vide cooking is the process of vacuum-sealing raw food in plastic pouches and cooking it slowly in a temperature-controlled water bath.} & \relzero \\
\midrule
\relzerop{K} & {\small Learn more sous vide at http://www.sousvidesupreme.com-----WHAT IS SOUS VIDE? -----The sous vide technique has been the secret of great chefs worldwide for decades. The SousVide Supreme is an amazing new all-in-one sous vide water oven designed to bring the sous vide cooking technique into home kitchens and restaurants at an affordable price.} & \relzero \\
\midrule
\relthreep{L} & {\small All kinds! Any type of meat—such as beef, pork, lamb, game, or poultry—is ideal for sous vide. It works especially well with fish and seafood, ensuring that these delicate foods are not overcooked. Almost any vegetable can also be cooked sous vide with delicious results, as can eggs and many fruits. You can even use it to make custard-style ice cream base, béarnaise sauce, crème Anglaise, custards, cheese, yogurt, and even cakes.} & \relthree \\
\midrule
\relonep{M} & {\small Sous Vide help - why did my salmon dry out at 120? Updated 1 month ago | 8. Sous Vide Salmon; 3 What temperature do I cook these foods on? Updated 3 months ago | 1. Chicken Breast Cookbooks Steak Beginner Cook Scrambling; 4 Potato encrusted salmon. Updated 1 month ago | 1. Salmon Potatoes Recipe Fixes; 5 What temperature do you have the grill on when cooking steak?} & \relone \\
\midrule
\relzerop{N} & {\small Here’s our answer: Water used to cook food sous vide is still clean and sanitary (provided your bag doesn’t break or leak), so you can use it to wash dishes, water your plants, hydrate your dog, bathe your baby, or fill your swimming pool. You can also use that water to cook sous vide several times, if you have the space to keep it.} & \relzero \\
\midrule
\relzerop{O} & {\small With the Anova, sous vide cooking is simple. All you need to get started is a pot or large container to hold water, heavy duty bags, a few clips, and your Anova Precision Cooker, of course! One of the best parts of sous vide cooking is that you don’t need to do much to your food before cooking.} & \relzero \\
\bottomrule
\end{tabular}
\caption{
Top 15 relevant passages for query \texttt{what types of food can you cook sous vide} from BM25 (Part 2 of 2).
More details in Table \ref{tab:bm25-passage-details-1}.
}
\label{tab:bm25-passage-details-2}
\end{table}

\begin{table}[]
\centering
\begin{tabular}{ll}
\toprule
Ranking Method & Ranking List  \\
\midrule
BM25 & \relzerop{A} $\succ$ \relthreep{B} $\succ$ \reltwop{C} $\succ$ \relzerop{D} $\succ$ \relzerop{E} $\succ$ \relthreep{F} $\succ$ \relzerop{G} $\succ$ \relzerop{H} $\succ$ \relzerop{I} $\succ$ \relzerop{J} $\succ$ \relzerop{K} $\succ$ \relthreep{L} $\succ$ \relonep{M} $\succ$ \relzerop{N} $\succ$ \relzerop{O}  \\
\midrule
GPT-3.5-Turbo & \relthreep{L} $\succ$ \relthreep{B} $\succ$ \relzerop{I} $\succ$ \relzerop{D} $\succ$ \relzerop{J} $\succ$ \relzerop{A} $\succ$ \reltwop{C} $\succ$ \relzerop{G} $\succ$ \relzerop{H} $\succ$ \relthreep{F} $\succ$ \relzerop{O} $\succ$ \relzerop{E} $\succ$ \relzerop{K} $\succ$ \relonep{M} $\succ$ \relzerop{N}  \\
\midrule
GPT-4 &  \relthreep{L} $\succ$ \relthreep{B} $\succ$ \relzerop{D} $\succ$ \relthreep{F} $\succ$ \relzerop{I} $\succ$ \relzerop{J} $\succ$ \reltwop{C} $\succ$ \relzerop{H} $\succ$ \relzerop{G} $\succ$ \relzerop{O} $\succ$ \relzerop{A} $\succ$ \relzerop{E} $\succ$ \relonep{M} $\succ$ \relzerop{N} $\succ$ \relzerop{K} \\
\midrule
Llama-3-70B &  \relthreep{L} $\succ$ \relthreep{B} $\succ$ \relthreep{F} $\succ$ \relzerop{I} $\succ$ \relzerop{A} $\succ$ \relonep{M} $\succ$ \relzerop{D} $\succ$ \relzerop{J} $\succ$ \relzerop{H} $\succ$ \relzerop{O} $\succ$ \reltwop{C} $\succ$ \relzerop{E} $\succ$ \relzerop{K} $\succ$ \relzerop{G} $\succ$ \relzerop{N} \\
\midrule
\rankfusion &  \relthreep{L} $\succ$ \relthreep{B} $\succ$ \relzerop{I} $\succ$ \relzerop{D} $\succ$ \relthreep{F} $\succ$ \relzerop{J} $\succ$ \relzerop{A} $\succ$ \reltwop{C} $\succ$ \relzerop{H} $\succ$ \relzerop{G} $\succ$ \relzerop{O} $\succ$ \relonep{M} $\succ$ \relzerop{E} $\succ$ \relzerop{K} $\succ$ \relzerop{N} \\
\bottomrule
\end{tabular}
\caption{
Comparisons of the rankings from different LLMs and aggregation.
The rankings are produced by LLM-based pairwise ranker using bubblesort.
\textbf{We include ICL and calibration in each individual LLM-based ranker}.
}
\label{tab:rank-passage-demo}
\end{table}

\subsection{Implementation details} \label{sec:implementation}

We use LLMs with a variety of sizes in our experiments: \textbf{GPT}~\cite{openai2023gpt}: GPT-3.5-Turbo-1106, GPT-4-1106; 
\textbf{LLaMA-3}~\cite{touvron2023llama}: LLaMA-3-70B, LLaMA-3-8B;
\textbf{Vicuna}~\cite{zheng2023judging,vicuna2023}: Vicuna-v1.5-13b;
\textbf{Mixtral}~\cite{jiang2024mixtral}: Mixtral-8x7b-v0.1;
\textbf{Gemma}~\cite{team2024gemma} Gemma-7B;
\textbf{Flan-T5}~\cite{chung2024scaling,tay2022ul2} Flan-UL2, Flan-T5-XXL.
This aims to explore the capacity of different sizes of LLMs and the trade-off between efficiency and ranking quality.

The LLM inference is implemented based on HuggingFace.
The instruction fine-tuned version of the model is used if available.
The temperature of LLM is set to $0$, which means $argmax$ will be applied to the candidate tokens during generation.
We use 2 $\times$ NVIDIA H100 80GB HBM3 and 4 $\times$ Tesla V100-SXM3-32GB GPUs to run our experiment.

\subsection{Prompt} \label{sec:prompt}

We show a comparison prompt example in Table \ref{tab:promp-example}.
The first two rounds of chat is the in-context Learning (ICL) prompt.
We expect the LLM can learn the order-agnostic property from the demonstration of ICL prompt.

\begin{table}[]
\centering
\begin{tabular}{l|p{0.8\linewidth}}
\toprule
User & {\small Given a query "{\color{violet} anthropological definition of environment}", which of the following two passages is more relevant to the query?

Passage A: "{\color{teal} Forensic anthropology is the application of the science of physical anthropology and human osteology in a legal setting, most often in criminal cases where the victim\'s remains are in the advanced stages of decomposition.nvironmental anthropology is a sub-specialty within the field of anthropology that takes an active role in examining the relationships between humans and their environment across space and time.}"

Passage B: "{\color{blue} Graduate Study in Anthropology. The graduate program in biological anthropology at CU Boulder offers training in several areas, including primatology, human biology, and paleoanthropology. We share an interest in human ecology, the broad integrative area of anthropology that focuses on the interactions of culture, biology and the environment.}"

Output Passage A or Passage B:} \\
\midrule
Assistant & {\small Passage: A} \\
\midrule
User & {\small Given a query "{\color{violet} anthropological definition of environment}", which of the following two passages is more relevant to the query?

Passage A: "{\color{blue} Graduate Study in Anthropology. The graduate program in biological anthropology at CU Boulder offers training in several areas, including primatology, human biology, and paleoanthropology. We share an interest in human ecology, the broad integrative area of anthropology that focuses on the interactions of culture, biology and the environment.}"

Passage B: "{\color{teal} Forensic anthropology is the application of the science of physical anthropology and human osteology in a legal setting, most often in criminal cases where the victim\'s remains are in the advanced stages of decomposition.nvironmental anthropology is a sub-specialty within the field of anthropology that takes an active role in examining the relationships between humans and their environment across space and time.}"

Output Passage A or Passage B:} \\
\midrule
Assistant & {\small Passage: B} \\
\midrule
User & {\small Given a query "\underline{what types of food can you cook sous vide}", which of the following two passages is more relevant to the query?

Passage A: "What is a Sous Vide Cooker? We said it before and we’ll say it again: a sous vide machine is a better way to cook. When you use a sous vide at home, you’ll be cooking with water. It’s not at all what you’re thinking. Your steak and chicken – any food really – will be placed in BPA-free, sturdy plastic bags and cooked."

Passage B: "Often, however, when you prepare food sous vide you’re packaging the food in plastic. (Often but not always. Eggs come in their own wrappers—their shells—and we can sous vide foods that set as they cook, like custard, yogurt, and chicken liver pâté, in glass canning jars.) "

Output Passage A or Passage B:} \\
\midrule
Assistant & {\small Passage: } \\
\bottomrule
\end{tabular}
\caption{
An example of prompt for a pairwise comparison with ICL.
For open-source LLMs, we can explicitly set the start of the last assistant's message to begin with "Passage:".
So that the LLM will generate the next token from "A" or "B".
}
\label{tab:promp-example}
\end{table}

\subsection{Performance}

The average number of comparisons required to rank a list is $3574.21 \pm 501.23$ for Bubblesort and  $972.77 \pm 47.69$ for Heapsort.
We benchmark the comparison rate of different LLM rankers in Table \ref{tab:bench}.
The prompt is longer after applying ICL, which decreases the LLM inference performance.
We only generate 1 token for each comparison on those open-source LLMs.
Each pairwise comparison involves prompting the LLM in two different passage permutations.
It requires 2 prompt and generate operations to finish a single pairwise comparison.
We benchmark the performance on NVIDIA H100 80GB HBM3 GPU.

\begin{table}
\centering
\begin{tabular}{ccl}
\toprule
Model & w/o ICL Rate & w/ ICL Rate\\
\midrule
Llama-3-8B & 31.14 & 20.94 \\
Llama-3-70B* & 6.41 & 2.92 \\
Mixtral-8x7B* & 10.33 & 7.44 \\
Vicuna-13B & 22.76 & 10.62 \\
Gemma-7B & 22.48 & 21.00 \\
Flan-UL2 & 11.77 & 5.00 \\
Flan-T5-XXL & 19.82 & 6.77 \\
\bottomrule
\end{tabular}
\caption{
Benchmark the comparison performance of different LLMs.
The \textbf{Rate} in the table means \textbf{\# comparisons per second}, each comparison include two query requests to the LLM.
*The Llama-3-70B and Mixtral-8x7B require 2 GPUs to run, other models run on single GPU.
}
\label{tab:bench}
\end{table}

\subsection{Analysis of Prompt Design Sensitivity} \label{sec:prompt_sensitivity}

\begin{table}[h]
\centering
\begin{tabular}{lcccc}
\toprule
Model & Baseline & ICL Only & Calibration Only & ICL + Calibration \\
\midrule
Bubblesort & 62.68±3.07 & 67.31±1.73 & 69.59±1.89 & \textbf{70.84±1.61} \\
Heapsort & 63.82±4.63 & 69.45±1.92 & 68.30±2.24 & \textbf{70.10±1.92} \\
\bottomrule
\end{tabular}

\caption{NDCG@10 scores (mean ± std) across different prompt designs}
\label{tab:prompt-sensitivity}
\end{table}

To evaluate the robustness of our proposed methods against variations in prompt design, we conducted additional experiments using GPT-4 to generate 10 different pairwise ranking prompt templates. Each template maintained the core task of comparing passage relevance while varying factors such as wording, formatting, and instruction style. We tested these prompts using Llama-3-8B on the TREC-DL2019 dataset under different configurations. The results in Table \ref{tab:prompt-sensitivity} demonstrate two key benefits of our approach:

\begin{enumerate}
\item \textbf{Enhanced Performance:} The combination of ICL and calibration substantially improves ranking quality, with NDCG scores increasing by approximately 8 points compared to the baseline across both sorting algorithms.
\item \textbf{Reduced Variance:} The baseline exhibits considerable sensitivity to prompt design, as evidenced by the high standard deviations (3.07-4.63). In contrast, configurations using ICL and calibration show markedly lower variance (1.61-1.92), indicating more stable performance across different prompt designs.
\end{enumerate}

\end{document}